\documentstyle[amssymb,preprint,eqsecnum,aps,prb,amstex]{revtex}

\begin{document}
\title{Magnetic contributions to the low-temperature specific heat of the
ferromagnetic insulator Pr$_{0.8}$Ca$_{0.2}$MnO$_3$}
\author{A. Wahl, V. Hardy, C. Martin, Ch. Simon}
\address{Laboratoire CRISMAT, Unit\'{e} Mixte de Recherches 6508, Institut des\\
Sciences de la Mati\`{e}re et du Rayonnement - Universit\'{e} de Caen, 6\\
Boulevard du\\
Mar\'{e}chal Juin, 14050 Caen Cedex, France.}
\date{\today}
\maketitle

\begin{abstract}
The $\Pr_{1-x}$Ca$_x$MnO$_3$ system exhibits a ferromagnetic insulating
state for the composition range $x\leq 0.25$. A metallic ferromagnetic state
is never realized because of the low hole concentration and the very small
averaged A-site cation radius. In the present study, the nature of the
magnetic excitations at low temperature has been investigated by specific
heat measurements on a $\Pr_{0.8}$Ca$_{0.2}$MnO$_3$ single crystal. The
decrease of the specific heat under magnetic field is qualitatively
consistent with a suppression of ferromagnetic spin waves in a magnetic
field. However, at low temperature, the qualitative agreement with the
ferromagnetic spin waves picture is poor. It appears that the large
reduction of the specific heat  due to the spin waves is compensated by a
Schottky-like contribution possibly arising from a Zeeman splitting of the
ground state multiplet of the Pr$^{3+}$ ions.
\end{abstract}


Hole-doped perovskite manganese oxides R$_{1-x}$AE$_x$MnO$_3$ (R and AE,
being trivalent rare-earth and divalent ions, respectively) are associated
with a wide variety of electronic and magnetic properties depending on the
value of $x$ and the averaged $A$-site cation radius, $\langle r_A\rangle $.%
\cite{REVIEW} These materials have recently been the subject of intense
studies due to intriguing phenomena such as charge/orbital ordering (CO)\cite
{CO} or colossal magnetoresistance (CMR).\cite{CMR} The latter is usually
interpreted by means of the double-exchange interaction (DE) theory.\cite{DE}
Although the DE mechanism cannot account alone for the temperature
dependence of the resistivity - recent theoretical works have claimed that
additional interactions, such as strong dynamical Jahn-Teller based
electron-lattice coupling are necessary to explain the magnitude of the
resistivity drop associated with the onset of ferromagnetism\cite{MIL95} -
such a scenario gives an interesting qualitative interpretation of coupled
ferromagnetic ordering and metallicity. Within such a framework, the
ferromagnetic (FM) ordering is related to a large electronic itinerancy i.e.
metallic behavior.

Among the hole-doped perovskite manganese oxides, the $\Pr_{1-x}$Ca$_x$MnO$%
_3 $ system (PCMO) is of great interest.\cite
{SMO00,LEE95,LEE99,TOM96,MAR99,MAI97,JIR85} For $0.3\leq x<0.8$, charge
ordering of $Mn^{3+}$and $Mn^{4+}$ (CO) is found and an antiferromagnetic
(AFM) ordering can be observed with N\'{e}el temperature ranging from 100K
to 170K for $x=0.8$ and $0.3$, respectively. A metallic state is never
realized under zero field for this composition range except upon application
of pressure,\cite{MOR97} light,\cite{COX98} high current\cite{ASA97} and
magnetic field.\ For $x\leq 0.25$, a ferromagnetic insulator (FMI) state is
always observed. Indeed, a ferromagnetic metallic state (FMM) can never be
realized because of the low hole concentration and the small averaged $A$%
-site cation radius, $\langle r_A\rangle $\cite
{MAR99,MAI97,JIR85,DAM97,ROD96,YAS00,HWA95} which result in a decrease of
the magnitude of W and, consequently, in a reduction of the effectiveness of
the DE interaction, mostly responsible for the absence of the FMM state.

We have performed a calorimetric study of the FMI $\Pr_{0.8}$Ca$_{0.2}$MnO$%
_3.$ The aim of this paper is to answer to the following question : What is
the nature of the magnetic contribution in the FMI state ? We expect the
analysis of the low temperature specific heat data $(C)$ to provide accurate
values of lattice, electronic and magnetic components in the FM state.\
According to both insulating and ferromagnetic behaviors reported for this
compound, one might expect the determination of the relevant contributions
to $C$ to be rather simple.\ However, as often reported in the literature,
we have found that an analysis based on a single set of zero field data
cannot yield unambiguous results. Thus, investigation of the magnetic field
induced specific heat change appears to be an essential tool.

In this paper, we focus on the reduction of the specific heat under magnetic
field. In order to shed further light on wether this reduction of specific
heat can be totally ascribed to a FMSW term, we have collected and analysed
specific heat data in different magnetic fields. There is a clear decrease
of the signal that is consistent with a suppression of the FMSW in a
magnetic field. Through a scaling approach, a quantitative agreement is
found with the FMSW theory for high temperature whereas the agreement is
poor at lower temperature. Besides, a modelling of the temperature and field
dependence of the low temperature magnetization within the sole FMSW picture
is also mainly unsuccessful. To fully describe the low temperature specific
heat, we suggest that an additional Schottky-like contribution might be
considered. Finally, the change of specific heat under magnetic field has
finally been modelled by a large reduction of specific heat due to FMSW that
is compensated by an increase in $C$ due to Zeeman splitting of the Pr$^{3+}$
ions. 

Using the floating-zone method with a feeding rod of nominal composition Pr$%
_{0.8}$Ca$_{0.2}$MnO$_3$, a several-cm-long single crystal was grown in a
mirror furnace. Two samples were cut out of the central part of this
crystal, one of them for resistivity measurements and the other for
magnetization and specific heat measurements. X-ray diffraction and electron
diffraction studies, which were performed on pieces coming from the same
part of the crystal, attested that the samples are single phased, and well
crystallized. The cell is orthorhombic with a Pnma space group, in agreement
with previously reported structural data. The energy dispersive spectroscopy
analyses confirm that the composition is homogeneous and close to the
nominal one, in the limit of the accuracy of the technique. The electron
diffraction characterization was also carried out versus temperature, from
room temperature to 92K. The reconstruction of the reciprocal space showed
that the cell parameters and symmetry remain unchanged in the whole domain
of temperature and, more especially, no extra reflections have been
detected. This electron diffraction observation, coupled with lattice
imaging, shows that, in our sample there is no charge ordering effect, even
at short range distances. All X-ray and electron diffraction observations
agree with previous published results for compounds of the same system.\cite
{HER99}

Specific heat measurements were carried out by using the two-$\tau $
relaxation method, at temperatures from 2.2 to 300 K and under magnetic
fields up to 9 T. The background signal, including the exact amount of
Apiezon N used to paste the sample on the platform, was recorded in a first
run, and it was then subtracted from the total heat capacity. The absence of
any significant field dependence of this background signal was carefully
checked in the low temperature range (2.2-15 K).

Resistivity measurements were performed by the standard four-probes
technique, at temperatures from 5 to 300 K, and fields up to 9 T.
Magnetization measurements were recorded by using a superconducting quantum
interference device magnetometer, at temperatures from 5 to 300 K, and
fields limited to 5 T. The $ac$ susceptibility was measured in zero $dc$
field with an alternating field $hac=1$ Oe for various frequencies.\smallskip

\smallskip

The temperature dependence of the specific heat ($C$), the $ac$
susceptibility ($\chi $) and the resistivity ($\rho $), up to 300 K and
under zero field, are shown in Figure 1 for our Pr$_{0.8}$Ca$_{0.2}$MnO$_3$
crystal. The data are shown up to high temperature to characterize the
overall behaviour of this compound. In all the measurements, the temperature
dependence is dominated by a salient feature at the
paramagnetic-ferromagnetic (PM-FM) transition, $T_c\sim 135$ K. The
transition in $C/T$ vs. $T$, corresponding to the onset of the FM ordering,
is evidenced as a sharp asymmetric anomaly [Figure 1(a)].\ Figure 1(b) shows
the temperature dependence of the real part of the susceptibility, $\chi
^{^{\prime }}$ ($hac=1\ $Oe, $1000$ Hz$)$. The rise in $\chi ^{^{\prime
}}(T) $ at a temperature around 135 K, associated with the onset of the FM
ordering is followed, on lowering the temperature, by a broad shoulder. One
observes a continuous decrease of $\chi ^{^{\prime }}(T)$ down to 50 K where
a trend to saturation occurs. The zero field temperature dependence of the
resistivity [Figure 1(c)] does not show any temperature-induced I-M
transition. However, a slight change in the slope takes place at the same
temperature as the other features described above. Isothermal $M$ vs. $H$
curve measured at 5 K [Figure 2], shows that magnetization first increases
up to 2.5 Tesla where the saturation value $(4.3$ $\mu _B$ $/$ f.u.$)$ is
reached. This is higher than the expected magnetic moment from Mn spins
contributions taking into account the relative concentration of Mn$^{3+}$
and Mn$^{4+}$ in the compound $(3.8$ $\mu _B$ / f.u.$).$ This feature is
often reported in the PCMO systems and is generally attributed to an
additional ferromagnetic moment arising from the FM ordering of the $\Pr $
spins with respect to the Mn spins.\cite{COX98,ROY00}

\smallskip

As a first step, we examine how informations concerning the relevant
contributions to $C$ can be extracted from zero field data. In this paper,
we confine the analysis to the data above 4 K for which the hyperfine term,
related to splitting induced by large local magnetic fields at the $\Pr $
and Mn nuclear spins, is assumed to be negligible. Various contributions are
usually considered for the specific heat in this low temperature range.
First, a phononic term arising from lattice vibrations has to be included.\
At low temperature, i.e. $T\leq 10$ K, this contribution can fairly be
approximated by a $\beta T^3$ term (a $T^5$ term may be added when an
analysis is carried out up to higher temperature). Although no carriers
contribution to $C$ is expected for $\Pr_{0.8}$Ca$_{0.2}$MnO$_3$ - this
contribution has the form of a linear temperature term, $\gamma T$, where $%
\gamma $ is proportional to the density of states at the Fermi level - many
insulator systems show the appearance of such a linear term in the specific
heat.\cite{SMO00,GHI99} If the gap is assumed to be zero then the magnetic
term associated with the FMSW excitations under zero field is $\delta
T^{3/2} $ with $\delta =0.113Ra^3(\frac{k_B}D)^{3/2}$ where $a$ is the
lattice parameter of the elementary perovskite cell and $R,$ the ideal gas
constant. The assumption of a zero spin gap is supported by numerous
experimental evidences.\cite{HAM96,LOF95}

In this section, we have attempted to model the zero field low temperature
data (4K-10K) assuming that the total specific heat might be comprised of
the 3 above terms. The fitting procedure including the 3 terms all together
does not converge at all if $\beta $, $\delta $ and $\gamma $ are left as
free parameters. This procedure yields unphysical values for $\delta $ and $%
\gamma $ depending on the temperature range investigated ; thus, as a first
step, we have considered only the FMSW contribution ($\delta T^{3/2})$
beside the lattice term$.$ Our result matches numerous reports where fitting
using the $\delta T^{3/2}$ term makes a very small contribution to specific
heat $(\delta =2$ mJ/K$^{5/2}$.mol$).$ \cite{SMO00,LEE99,HAM96,GHI98}
However, the observation of a magnetic contribution to the specific heat of
the FM phase of the doped manganite has generated conflicting results and
the question of whether a set of zero field data is sufficient to observe a
magnetic contribution was often raised.

A fitting of equally good quality can be obtained if one includes a sole
linear term beside the lattice contribution. As pointed out above, including
such a linear contribution is not senseless in our ferromagnetic insulator.
Such a feature has been reported for the first time in Eu$_x$Sr$_{1-x}$S
with $x=0.4$\cite{MAR80} and in CuMn alloys.\cite{MES80} Nevertheless, the
linear coefficient $\gamma $ is found to be $4$ mJ/K$^2$.mol and such a
small value is not consistent with values usually associated with spin
disorder.\cite{SMO00,GHI99} It must be pointed out that, in both cases, the
fittings are achieved with $\beta _3$ parameters significantly larger than
those usually associated with lattice contributions in hole-dope manganites (%
$\theta _D<250K)$. However, although a softening of the lattice in the
insulating phase has already been reported in many papers\cite
{LEE99,WOO97,OKU00,OKU98}, this can not explain this very low Debye
temperature.

As reported in previous work,\cite{LEE99,ROY00,GHI99,HAM96,GHI98,WOO97,OKU00}
we were unable to obtain reliable results concerning a FMSW contribution
and/or a linear term characterizing the disorder using a standard fitting
procedure for zero field data. Indeed, it is obvious that the present low
temperature specific heat measurements without field cannot allow us to
determine a unique set of values for the contributions presumably involved
in the analysis. Thus, we expect the study of the change of specific heat
under magnetic field to shed light on the nature of the contributions that
occur at low temperatures.

The curves for 0, 3, 5, 7 and 9 Tesla are shown in Figure 3. These data were
registered upon warming the sample which was previously cooled under field
(FCW mode). Measurements were also carried out with cooling under zero field
(ZFC mode) and no significant change was observed. A decrease of the
specific heat is observed under magnetic field ; this feature is more
obvious in the inset of Figure 3 where the field dependence of the specific
heat is directly measured at 10 K after zero field cooling.\ Such a decrease
of the specific heat under magnetic field is at first sight consistent with
thermodynamic expectations for a ferromagnet $\left( \frac{\partial C}{%
\partial H}\prec 0\right) $. The results obtained for $\Pr_{0.8}$Ca$_{0.2}$%
MnO$_3$ are in {\it qualitative} agreement with the suppression of the
thermal excitations of the FMSW in the presence of a field induced gap.
However, it is essential to check if the specific heat decrease under
magnetic field is in {\it quantitative} agreement with the FMSW theory.

Under magnetic field, in the FM state, the magnetic term is expressed in the
following way : 
\begin{equation}
C_{FMSW}(H,T)=Ra^3\left( \frac{k_BT}D\right) ^{3/2}\ F\left( \frac HT\right)
\quad where\quad F\left( \frac HT\right) =\frac 1{4\pi ^2}\int\limits_{\frac{%
g\mu _BH}{k_BT}}^\infty \frac{x^2e^x}{\left( e^x-1\right) ^2}\sqrt{x-\frac{%
g\mu _BH}{k_BT}}\ dx
\end{equation}

In this expression, the FMSW excitation at zero field is assumed to show no
gap and $F\left( 0\right) =0.113$.

Hence, $\Delta C_{FMSW}=C_{FMSW}(0,T)-C_{FMSW}(H,T)$ is written in the form
: 
\begin{equation}
\Delta C_{FMSW}=0.113Ra^3\left( \frac{k_BT}D\right) ^{3/2}-Ra^3\left( \frac{%
k_BT}D\right) ^{3/2}F\left( \frac HT\right) =T^{3/2}G\left( \frac HT\right)
\end{equation}
\begin{equation}
with\quad G\left( \frac HT\right) =Ra^3\left( \frac{k_B}D\right)
^{3/2}\left[ 0.113-F\left( \frac HT\right) \right]
\end{equation}

To avoid any experimental problem arising from the choice of a fitting
procedure, the scaling approach seems to be the most promising way to
emphasize the FMSW contribution. According to the above expression, $\frac{%
\Delta C_{FMSW}}{T^{3/2}}$ only depends on $\frac HT$ ; this could provide
us a direct test for the reliability of the standard FMSW model. If all the
magnetic specific heat arises from the field induced suppression of the
FMSW, the plot $\frac{\Delta C_{data}}{T^{3/2}}$ $vs$ $\frac HT$ should show
the same trend for data recorded for different temperatures and under
various magnetic fields. Figure 4 illustrates that, at low temperature
(highest values of $\frac HT$), $\Delta C_{data}=C_{data}(T,0)-C_{data}(T,H)$
cannot be fully described by considering a magnetic contribution arising
from the {\it sole} FMSW. However, while T is increased (for lower $\frac HT$
values), a superimposition of the data is observed suggesting that the
description in term of FMSW becomes to be relevant. The solid line in Figure
4 corresponds to the calculated $G\left( \frac HT\right) $ scaling function
considering $D\approx 15\pm 3\ meV.\stackrel{\circ }{A}^2$. Considering this
latter value of the FMSW stiffness, the reduction of the specific heat as a
function of the applied field is also compared to the prediction of the FMSW
theory in Figure 5 for a temperature of 13 K $\left( 0\leq \frac HT\leq
0.69\right) $. The value of the FMSW stiffness might seem very small
compared to the ones derived through neutron scattering experiments for
other manganites.\cite{LYN96,LYN97,END97} However, as emphasized by Roy et $%
al.$\cite{ROY00}, the very soft SW in $\Pr_{0.8}$Ca$_{0.2}$MnO$_3$ appears
to be a general feature of the low doped region of the PCMO. These authors
invoke a drastic effect of the ferromagnetic moment associated with the $\Pr 
$ ions.

Before going further, a modelling of the H- and T-dependence of the
magnetization within the FMSW framework could support the above observation.
The ''Bloch $T^{3/2}$ law'', in which the presence of the field has been
properly taken into account using the standard FMSW picture, has been
tested. According to the procedure proposed by Smolyaninova {\it et al.}\cite
{SMO97}, the low temperature magnetization follows the form : 
\begin{equation}
M(0,H)-M(T,H)\varpropto T^\alpha
\end{equation}

where $M(0,H)$ is the extrapolation of $M(T,H)$ back to $T=0$. $M(0,H)$ for
fields 3, 4 and 5 Tesla agrees with the saturation value $(\approx 4.3\mu
_B/f.u.)$ derived from isothermal $M$ $vs$ $H$ measured at 5 K.\ The power
law fits rather well the measured low temperature magnetization (within the
range 5K - 15K) with $\alpha =$2.7, 3.4 and 3.35 for 3, 4 and 5 Tesla,
respectively. Using the standard SW picture, the magnetization per unit
volume at low temperature is given :

\begin{equation}
M(0,H)-M(T,H)=g\mu _B\left( \frac{k_BT}{4\pi D}\right) ^{3/2}f_{3/2}\left[
g\mu _B\left( H-NM\right) /k_BT\right]
\end{equation}

where $f_p\left( y\right) =\sum\limits_{n=1}^\infty e^{-ny}/n^p$ and $NM$ is
the demagnetization field (0.4 Tesla). Over the investigated range of
temperature, the above relation can be well approximated by a power law (See
inset Figure 6). On Figure 6, the field dependence of $\alpha $ expected
from the FMSW picture is compared to the values derived from the data. As
seen in Figure 6, there exists a large discrepancy which indicates that the
low temperature magnetization does not show the behavior expected for a
simple spin wave picture.

\smallskip Let us now turn back to the specific heat data.\ As shown before,
for the very low temperature range, the change in the specific heat does not
follow the temperature and magnetic field dependences of the {\it sole}
FMSW. Indeed, it is obvious that the magnitude of the $\Delta C_{data}$ is
not important enough to be described in such a way : there exists in the
very low temperature range an excess of specific heat under magnetic field.\
Hence, we speculate that other contributions could also change under
magnetic field. Using the calculated contribution of the FMSW at low
temperature, we can obtain the field and temperature dependence of this
excess specific heat :

\begin{equation}
\Delta C_{excess}=\Delta C_{FMSW}-\Delta C_{data}
\end{equation}

\smallskip Results are illustrated in Figure 7. Bell-shaped curves are
obtained with maxima slightly shifted to higher temperature as the field
increases ; the magnitude of the excess specific heat is also observed to
increase with increasing fields. Such a behavior has a Schottky-like
appearance and we speculate that this magnetic Schottky anomaly occurs due
to the Zeeman splitting of the crystal field ground state multiplet of the $%
\Pr^{3+}$ ions\cite{ROS99} under magnetic field. Clearly, a true two-level
system exists for spin $\frac 1{2\text{ }}$ ions only ; for spins $J\neq 
\frac 12$, specific heat should be generalized to the multilevel Schottky
function on the basis of the Langevin theory. However, a simplified
two-levels system, where the effective moment of the $\Pr^{3+}$ in the
ground state is $\mu $ and $\Delta $, the size of the two levels system,
should be a rather good approximation for the observed effect in specific
heat. $\Delta $ is expressed in the form : $\Delta =2\mu \left(
H_{int}+H\right) $ where $H_{int}$ is the internal field at the Pr$^{3+}$
site and $H$, the applied magnetic field.

Hence, we have fitted the data $\left( \Delta C_{excess}\right) $ to a
two-level Schottky function :

\begin{equation}
\Delta C_{Sch}(T,H)=C_{Sch}(T,H)-C_{Sch}(T,H=0)
\end{equation}

\begin{equation}
where\ \ \ \ C_{Sch}(T,H)=n_{Sch}R\left( \frac \Delta {k_BT}\right) ^2\frac{%
\exp \left( \Delta /k_BT\right) }{\left( 1+\exp \left( \Delta /k_BT\right)
\right) ^2}
\end{equation}

\smallskip Fits for 3, 5 and 9 Tesla, shown as solid curves on Figure 7, are
in rather good agreement with the experimental data. The resulting energy
splitting (inset Figure 7) give $H_{int}$ = 22 Tesla and correspond to a
magnetic moment of approximately $0.3\mu _B$. This value is similar to $%
0.5\mu _B,$ estimated at low temperature from magnetization measurements and
neutron scattering experiments\cite{COX98} and $0.8\mu _B$ from high field
magnetization.\cite{Thomas}\ Thus, the large reduction of specific heat
under magnetic field is partially compensated by a Schottky-like
contribution arising from the ground state level splitting of the rare earth
element. The level splitting derived from the specific heat data is in rough
agreement with the one estimated from high field magnetization in $\Pr_{0.7}$%
Ca$_{0.3}$MnO$_3$ ($5K$ for $H=6T).$\cite{Thomas}

This low tempearture anomaly has no corresponding feature in the
magnetization except the fact that the latter can not be described within
the framework of the SW theory. Moreover, knowing the excess and spin wave
contribution it is then possible to determine the phonon background which
can be used as a reliable test for our description. This yields a Debye
temperature $\theta _D\sim 325K$ which is typical of values already reported
for the PCMO system.\cite{LEE99}

There have already been reports of specific heat data in related materials
in which Schottky like anomalies due to Zeeman splitting of the rare eath
state is clearly seen. However, we adress here the subtle role of two
interacting spin systems (Pr and Mn ones) on the low temperature specific
heat of a hole doped manganite. We speculate that the effect is the
following : the Pr moments experience an effective molecular field produced
by the Mn moments ($H_{int})$ and a concomitant modification of the specific
heat ($\Delta C_{excess})$ is obtained due to the molecular field induced
splitting of the crystal field ground state multiplet of the $\Pr^{3+}.$

In conclusion, low temperature specific heat measurements have been carried
out on the ferromagnetic insulator $\Pr_{0.8}$Ca$_{0.2}$MnO$_3$.
Measurements of the specific heat under magnetic field have allowed us to
obtain reliable informations on the nature of the magnetic excitations
involved in the low temperature thermodynamic of $\Pr_{0.8}$Ca$_{0.2}$MnO$_3$%
. We have shown that an analysis in terms of ferromagnetic spin waves is not
sufficient for describing the entire magnetic contribution to specific heat.
An additional contribution of Schottky-like nature, probably arising from
the interaction of the Mn and Pr spin sytems has to be included.

\section{Figure Captions}

\begin{description}
\item[FIG. 1]  Temperature dependence of (a) the specific heat divided by
temperature $C/T$ under zero field ; (b) the real part of $ac$
susceptibility $\chi ^{\prime }$ ($hac=1$ Oe$;1000$ Hz$)$ ; (c) the
normalized resistivity under zero field.

\item[FIG. 2]  Field dependence of the magnetization at 5K

\item[FIG.\ 3]  $C$ -vs.- $T$ curves under various fields, recorded in the
FCW mode [0 T ($\bigcirc $); 3 T ($\Box $); 5 T ($\Delta $); 7 T ($\nabla $%
); 9 T ($\Diamond $)]. Inset displays specific heat values at 10 K as a
function of the magnetic field

\item[FIG. 4]  $\frac{\Delta C_{data}}{T^{3/2}}$ $vs.$ $\frac HT$ for
different temperatures and fields. The solid line is the theoretical scaling
function $G\left( \frac HT\right) $given in the text. Several symbols stand
for the quantities under different magnetic fields.

\item[FIG. 5]  $\Delta C_{data}=C_{data}(0,T)-C_{data}(H,T)$ at 13 K as a
function of magnetic field. The solid line is the calculated FMSW
contribution.

\item[FIG. 6]  The solid line indicates the expected field dependence of the
exponent within the FMSW picture.\ The black square are from data. Insert :
Magnetization vs temperature for H=3T. The line is the best fit to the power
law (see text).

\item[FIG. 7]  $\Delta C_{excess}\ vs.\ T$ for fields 3, 5 and 9 Tesla.
Solid curves : fits to a two-level Schottky function. The inset shows the
resulting energy splitting as a function of the applied magnetic field.
\end{description}

\end{document}